\documentclass[5p,authoryear]{elsarticle}
\usepackage{amsmath}
\usepackage[caption=false]{subfig}
\usepackage{graphicx}
\usepackage{url}

\begin{document}

\title{Jumping the energetics queue: Modulation of pulsar signals by extraterrestrial
civilizations}

\author[1]{Jayanth Chennamangalam\corref{cor1}}
\ead{jayanth@astro.ox.ac.uk}
\author[2,3,4]{Andrew P. V. Siemion}
\author[1,5]{D. R. Lorimer}
\author[2]{Dan Werthimer}

\cortext[cor1]{Corresponding author. Present address: Astrophysics,
    University of Oxford, Denys Wilkinson Building, Keble Road, Oxford OX1 3RH, UK}
\address[1]{Department of Physics and Astronomy, West Virginia University, PO Box 6315, Morgantown, WV 26506, USA}
\address[2]{Spaces Sciences Laboratory, University of California, Berkeley, CA 94720, USA}
\address[3]{ASTRON, PO Box 2, 7990 AA Dwingeloo, The Netherlands}
\address[4]{Department of Astrophysics, Radboud University, PO Box 9010, 6500 GL Nijmegen, The Netherlands}
\address[5]{NRAO, Green Bank Observatory, PO Box 2, Green Bank, WV 24944, USA}

\begin{abstract}
It has been speculated that technological civilizations evolve along an energy
consumption scale first formulated by Kardashev, ranging from human-like civilizations
that consume energy at a rate of $\sim 10^{19}$ erg s$^{-1}$ to hypothetical highly
advanced civilizations that can consume $\sim 10^{44}$ erg s$^{-1}$. Since the transmission
power of a beacon a civilization can build depends on the energy it possesses, to make
it bright enough to be seen across the Galaxy would require high technological advancement.
In this paper, we discuss the possibility of a civilization using naturally-occurring radio
transmitters -- specifically, radio pulsars -- to overcome the Kardashev limit of
their developmental stage and transmit super-Kardashev power. This is achieved by the
use of a modulator situated around a pulsar, that modulates the pulsar signal,
encoding information onto its natural emission. We discuss a simple modulation model
using pulse nulling and considerations for detecting such a signal. We find that a
pulsar with a nulling modulator will exhibit an excess of thermal emission peaking in the
ultraviolet during its null phases, revealing the existence of a modulator.
\end{abstract}


\maketitle

\section{Introduction}

The Kardashev scale \citep{kar64} classifies civilizations according to their
ability to consume energy. The human civilization is the prototypical Kardashev
Type-I civilization\footnote{The Kardashev scale has been redefined and expanded by
others \citep[see, for example,][]{hor93}, but in this paper, we follow the original
definition presented in \citet{kar64}, with no consequence to our treatment.},
consuming energy at the rate of
$\sim 4 \times 10^{19}$ erg s$^{-1}$. A Kardashev Type-II civilization consumes
$\sim 4 \times 10^{33}$ erg s$^{-1}$ -- equivalent to the energy output of a Sun-like
star. A Type-III civilization would be capable of consuming
$\sim 4 \times 10^{44}$ erg s$^{-1}$, which is of the order of the
luminosity of galaxies. If an extraterrestrial intelligence (ETI)
decides to build a radio beacon to announce their presence in the Galaxy to
prospective listeners for the purpose of eventually establishing a communication
channel \citep{coc59}, such a radio beacon would necessarily have a transmission
power not more than what that civilization consumes. In this paper, we assume
that the transmission power of an ETI beacon is of
the same order of magnitude as their energy consumption. Following \citet{kar64}, we can
calculate the power required to isotropically transmit a signal with a
bandwidth $\Delta f$ across the Galaxy, such that it can be received at an
Arecibo-like radio telescope with a signal-to-noise ratio $S/N$, as
\begin{equation}
P \approx 6.6 \times 10^{24} \left(\frac{\Delta f}{\rm Hz}\right)
\left(\frac{S/N}{10}\right)~{\rm erg~s^{-1}}.
\end{equation}
Traditional radio SETI experiments search for narrow-band ($\sim$ 1 Hz) signals. Even
for such narrow-band signals, a detectable $S/N$ would imply a transmission power that can
be generated only by civilizations that are much more advanced than Type-I. The main
drawback of using narrow-band signals as beacons is
that the ETI is forced to choose some special frequency that may not be monitored by
potential receivers. The solution to this problem is to transmit over a larger
bandwidth, but since $P \propto \Delta f$, the power requirement increases. For instance,
for $\Delta f = 1$ GHz, $P \sim 10^{33}~{\rm erg~s^{-1}}$, which can only be produced by
civilizations that are at least Type-II. The power requirement can be reduced
by trading it off with the solid angle of transmission, but barring very
narrow beams, the civilization still needs to be fairly advanced to provide the
necessary power. For instance, transmission with 1 GHz of bandwidth using a 1 square arcmin. beam
would need $P \sim 10^{25}~{\rm erg~s^{-1}}$. Less advanced civilizations that
wish to maximize sky coverage without incurring higher costs, however, can work
around this problem by making use of appropriate naturally-occurring radio
transmitters. In this paper, we propose that an ETI that is moderately more
advanced than humans but not yet achieving a higher Kardashev type, may be able to
use radio pulsars as sources of power at levels otherwise unachievable, modulating the
broad-band pulsar signal for communication. The minimum requirement for such an endeavour
would only be the ability to build and launch a modulating satellite to a nearby pulsar.

A pulsar is a neutron star that emits coherent radio radiation from its magnetic
poles \citep[see][]{lor05}. Pulsars are fast-rotating, and usually detected due
to the fact that an offset exists between their magnetic and rotational axes,
causing them to appear as periodic signals, with an observer typically
receiving one pulse per one complete rotation of the pulsar. The radio
luminosity of a pulsar with spin period $P$ situated at a distance $d$ from an
observer is given in terms of the measured flux density as
\begin{equation}
L = \frac{4 \pi d^2}{\delta} \sin^2 \left(\frac{\rho}{2}\right) \int_{f_1}^{f_2} S_{\rm mean}(f)~df,
\end{equation}
where $\delta = W_{\rm eq} / P$ is the pulse duty cycle ($W_{\rm eq}$ is the
equivalent pulse width), $\rho$ is the radius of the pulsar emission cone,
the integrand is the mean flux density of the pulsar as a function of frequency
$f$, and $f_1$ and $f_2$ bound the spectral range of the observation. Using
typical values of $\delta$ and $\rho$, a pulsar with $P = 1$ s situated at a
distance of 1 kpc, with a measured 1400 MHz flux density of 1 mJy, would have a
radio luminosity $\approx 7.4 \times 10^{27}$ erg s$^{-1}$. On the Kardashev
scale, such a pulsar would therefore correspond to a beacon produced by a civilization
between Type-I and Type-II. We speculate that a civilization with the minimum capability
of sending a spacecraft to a nearby pulsar to install an orbital modulator for the
sweeping pulsar beam would be able to harness the energy emission of pulsars without
actually building and operating a transmitter so
powerful (or being capable of doing so).

Previous works have considered extraterrestrial civilizations making use of
naturally-occurring phenomena to announce their presence to any listeners. For
example, \citet{cor93} has suggested that extraterrestrial civilizations may
make use of astrophysical masers to amplify engineered signals, thereby
transmitting more power than their position on the Kardashev scale might allow
them to. A critical drawback of using a maser-based communication system is
that masers are usually directional, and hence require the transmitter and
receiver to be serendipitously aligned. Pulsar beams, on the other hand, albeit
directional, are swept around due to the rotation of the star, thereby covering
a much larger area of the sky, increasing the probability of detection. A
system that makes use of pulsars, in addition to being used as beacons, can
also be configured for directional communication, with say, a distant
spacecraft or planetary system. \citet{fab77} and \citet{cor97} have discussed
the possibility of generating X-ray pulses by dropping matter onto the surface of a
neutron star, or modulating the X-ray emission of accreting neutron stars.
\citet{lea08} has proposed that ETI may modulate the period of Cepheid variables to
achieve signaling, by triggering pulsations using neutrinos beamed to the stellar core.

\citet{cor95} and \citet{sul95} postulate that ETI would employ
`astrophysical coding' -- i.e., transmitting signals that can be detected using
astrophysical signal analysis -- in beacons. They argue that such a signal is
more likely to be detected because astronomers would be able to easily analyse
it. The idea proposed in this paper is a kind of astrophysical coding
technique and enjoy the benefit of higher likelihood of detectability.

The outline of this paper is as follows: In \S\ref{sec_modmech}, we describe our proposed
modulation mechanism, and in \S\ref{sec_infcon}, we discuss the
information content of the beacon. In \S\ref{sec_obs}, we discuss potential
observational signatures of artificial modulation, and in \S\ref{sec_disc}, we analyse
energy considerations for this signalling scheme, before concluding in \S\ref{sec_conc}.

\section{Modulation mechanism}\label{sec_modmech}

Installing a modulator on a pulsar would require considerations of the emission
geometry of the pulsar being engineered. If we assume an inclination angle
$\alpha = 90^\circ$ (i.e., the magnetic axis orthogonal to the spin axis),
the modulating satellite could orbit synchronously with the pulsar spin period
to allow the signal to be transmitted over the entire area of the sky covered
by the pulsar beam. In the more typical case of non-orthogonal axes, a polar
orbit in which the satellite intersects the pulsar beam periodically would
result in directional transmission. A scaffolding shell around the pulsar
in which modulating elements are placed at locations where the pulsar
beam intersects with the scaffold would result in the ability to cover the
entire beaming solid angle of the pulsar.

\begin{figure*}
\begin{center}
{\begin{minipage}[t]{0.45\linewidth}
        \vspace{0pt}
        \centering
        \includegraphics[scale=0.5]{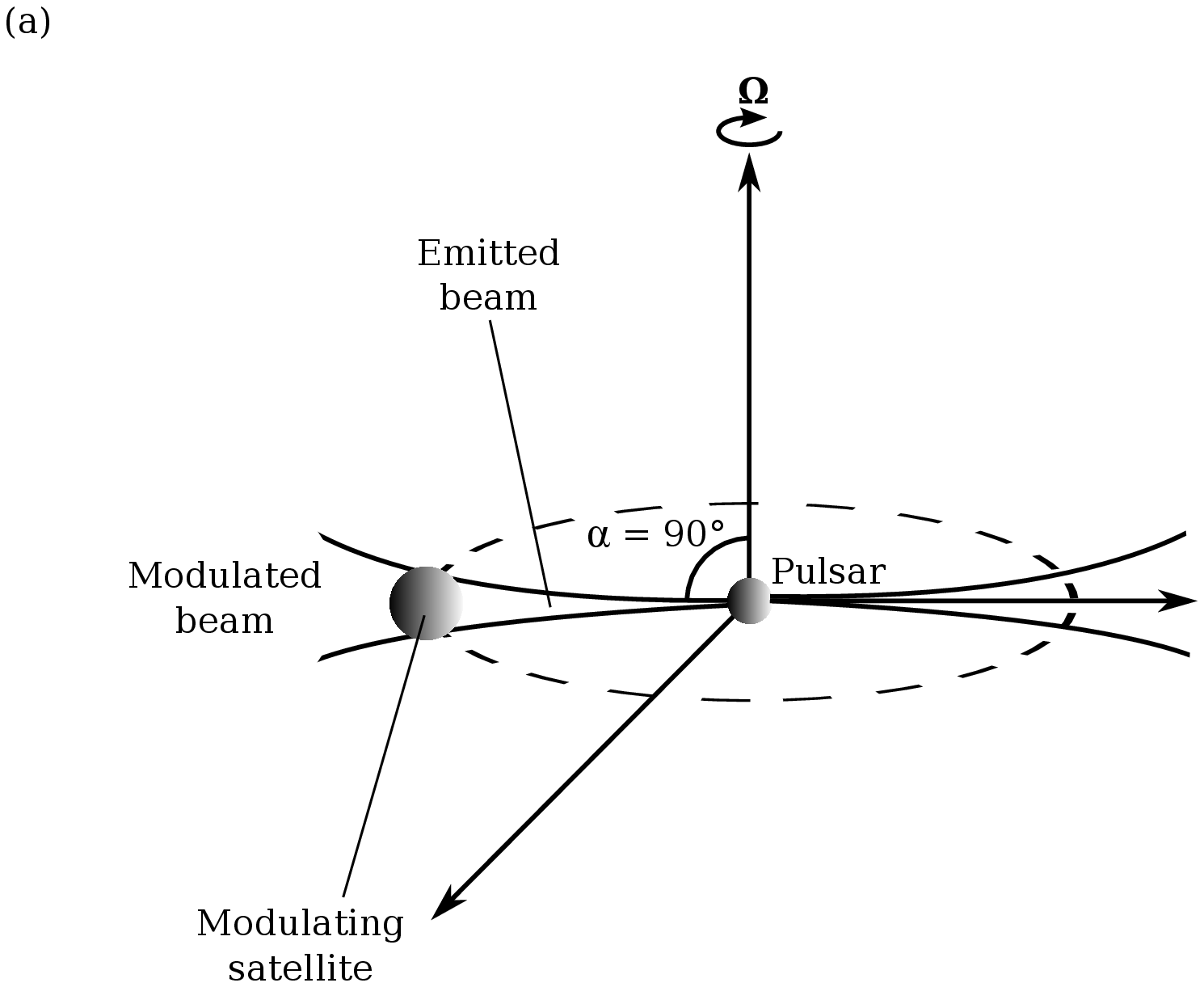}
      \end{minipage}}
{\begin{minipage}[t]{0.45\linewidth}
        \vspace{0pt}
        \centering
        \includegraphics[scale=0.5]{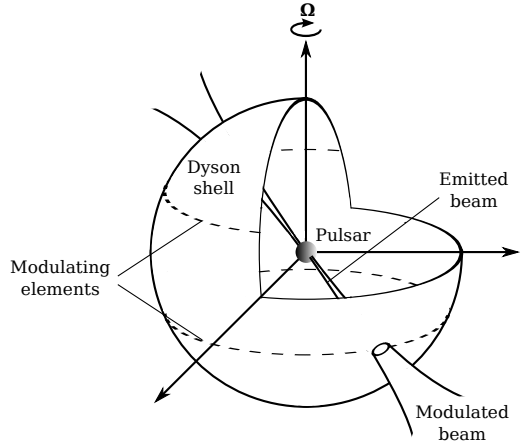}
      \end{minipage}}
\caption{(a) Schematic of a modulating satellite co-rotating with a pulsar that
has an inclination angle $\alpha = 90^\circ$; (b) Cross-section of a Dyson shell
around a pulsar. Modulating elements are placed along the loci of the pulsar beams on
the shell.
\label{fig_schematic}}
\end{center}
\end{figure*}

We first consider a toy model of an orbital modulator that is synchronous with
the pulsar rotation, assuming that the inclination angle of the pulsar beam,
$\alpha = 90^\circ$, as shown in Figure~\ref{fig_schematic}(a). For a pulsar with mass $M$ and period $P$, equating
centripetal acceleration to the acceleration due to gravity gives an orbital
radius
\begin{equation}
r \approx 1.7 \times 10^3 \left(\frac{M}{1.4M_\odot}\right)^{1/3} \left(\frac{P}{\rm s}\right)^{2/3} {\rm km}.
\end{equation}
For a canonical 1.4$M_\odot$ pulsar with $P = 1$ s, this gives
$r \approx 1700$ km, with a tangential velocity component of approximately 4\%
the speed of light. To probe the structural integrity of the satellite at
this distance, we model the satellite as a solid steel cylindrical bar 10 m in length
and 1 m in radius, oriented in such a way that the long axis is directed radially
outwards from the pulsar. The elongation of the bar due to the differential gravity on
either of its ends is of the order of 10$^{-5}$ m, and therefore, is inconsequential.

Instead of a satellite, a civilization capable of
advanced astronomical engineering could build an equatorial ring around the pulsar that
covers the entire area swept by the beam. A less desirable option would be to have a
satellite in a non-synchronous orbit periodically intercepting the pulsar beam, but this
would severely reduce the beaming fraction of the modulated beam, and also make message
reconstruction more difficult.

The typical case of non-orthogonal beams, however, is more complicated. A
modulating satellite in a polar orbit that intercepts the pulsar beams periodically could
be built, but this has the problem of low beaming fraction, which would not
serve as a beacon, but could be used for directional communication. For a
beacon, the last option -- albeit one that would require a significant amount
of astronomical engineering -- would be a scaffold around a pulsar akin to a
Dyson sphere \citep{dys60}, with modulating elements placed at the points where
the pulsar beam intersects the scaffold, as shown in Figure~\ref{fig_schematic}(b).
Traditional Dyson structures around main-sequence stars provide
general-purpose energy for the consumption of an advanced civilization.
In the case of a Dyson sphere around a pulsar as
outlined in this paper, the energy of the host star is used only for
producing a beacon. The materials that make up the scaffold, and the
structural engineering of the scaffold should be such that it should not interact with the
particles and field lines within the pulsar magnetosphere, except at the modulating
elements. The radius of the shell should be large enough such that
the pulsar emission region lies within this shell. \citet{kij03} give a
semi-empirical formula to calculate the
heights of emission regions of pulsars, which, for a 10 km-radius pulsar, is
\begin{equation}
r_{\rm em} \approx 400 \left(\frac{f}{\rm GHz}\right)^{-0.26}
\left(\frac{\dot{P}}{10^{-15}}\right)^{0.07}
\left(\frac{P}{\rm s}\right)^{0.30} {\rm km},
\end{equation}
where $f$ is the frequency of radio emission
at $r_{\rm em}$, $\dot{P}$ is the period derivative, and $P$ is the period
of the pulsar. Assuming that the lowest frequency of interest to the modulation
is 10 MHz, a pulsar with $P = 1$ s and
$\dot{P} = 10^{-15}$ s s$^{-1}$ would have a maximum emission height of
interest of $\sim 1300$ km. This gives the minimum radius of the Dyson
sphere.

As in any communication system, the modulation could be one of many types. It
could be amplitude modulation, frequency modulation, or phase modulation,
either analogue or digital. In this paper, we consider the simplest case, where
an amplitude modulator toggles between 0\% modulation (modulator transparent to
pulsar radiation) and 100\% modulation (modulator opaque to radiation) to
achieve preferential nulling, resulting in single-bit data transmission. For
the sake of simplicity, we also assume that the nulling is
frequency-independent, that is, during a null, the entire radio emission of the
pulsar is blocked. The information content of this system is discussed in \S\ref{sec_infcon}.

We do not speculate on the nature of the modulator as it would most
likely be based on technology not yet invented by humans, although it would
seem that the signal-modulating mechanism could be based on confined plasma, or
perhaps, electro-optic modulators \citep[see][]{pur10}. In the case of nulling
modulation, the modulating material would need to scatter, absorb or redirect the entire
radiation falling on it. If the radiation is absorbed, this would manifest as an
increase in the temperature of the modulator and could show up as thermal radiation
when the energy is re-radiated. This is discussed in more detail in \S\ref{sec_obs}.
The effect of radiation-induced heating of the modulator on the
lifetime of the system is discussed in \S\ref{sec_disc}.

\section{Information content}\label{sec_infcon}

A single-bit modulation system as mentioned in the previous section would
support only low bit rates. For the nulling method, the data rate,
$R = 1 / P$ bits per second, where $P$ is the spin period of
the pulsar in seconds. Even utilizing the fastest pulsars, the transmission
rate would be less than 1 Kbps. In this model, we have assumed that the nulling
is independent of frequency. A more complex system could make use of
frequency-dependent nulling, thereby increasing the data rate of the signal.
Further increase in data rate using amplitude modulation would require more
complex modulation mechanisms wherein the amplitude of a pulse varies over a
range of modulation depths. Another possibility is using a modulating signal
whose frequency is much larger than the pulsar period (but less than the
`carrier' frequency, or the radio frequency). This would manifest as narrow
features in the time domain, within the on-pulse of the pulsar.

The artificially-modulated pulsar signal contains another piece
of `information' that is astrophysically-coded -- the fact that the ETI
identified the neutron star that was engineered as a pulsar indicates that
their civilization may be based on at least one planet that is within the
beaming solid angle of the pulsar. If the inclination angle of the pulsar beam
can be determined, this helps derive a coarse constraint on the location of
the civilization.

\section{Observable effects}\label{sec_obs}

\begin{figure}
\includegraphics[width=\linewidth]{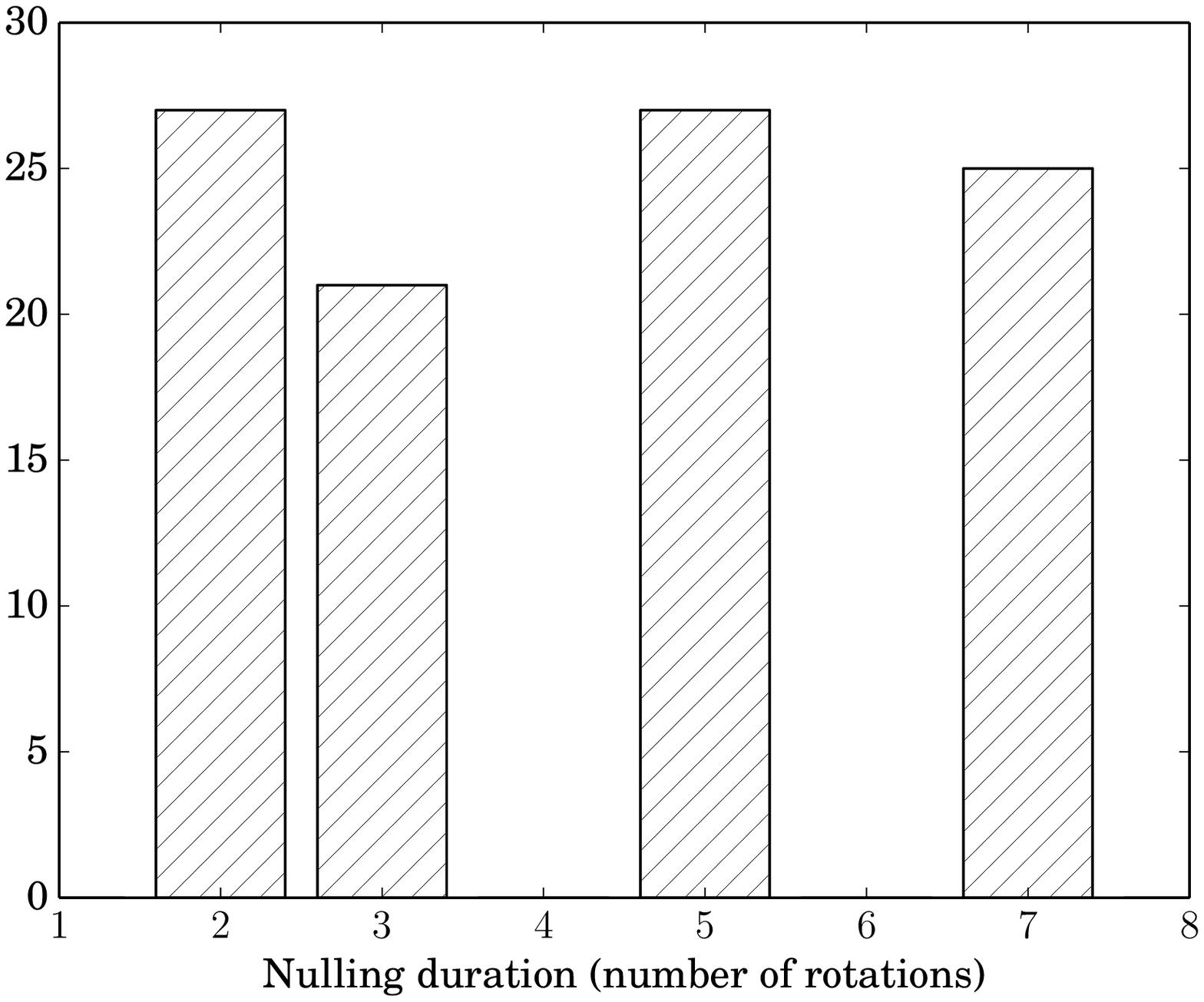}
\caption{Example histogram of null duration in terms of number of rotations of the
pulsar. This histogram is indicative of a non-natural nulling process.
\label{fig_nullhist}}
\end{figure}

Pulse nulling \citep{bac70} is observed in many pulsars, and is usually
attributed to changes in the plasma currents in the pulsar magnetosphere
\citep[see, for example,][]{wan07}, although this explanation has not been
conclusively established. Statistical studies of nulling and the possibly
related phenomena of mode-changing and drifting sub-pulses have the potential
to determine any sign of non-natural processes in action. \citet{red09} treated
the pulse-null stream for a set of pulsars as a binary sequence and performed a
statistical runs test, and found that nulling is not random in many pulsars
that exhibit the phenomenon. But as shown by \cite{cor13}, this just indicates
a different random Markov process in action. An artificially-nulled pulsar would also show up
as non-random in an analysis as done by \cite{red09}, but in general, it is unclear
how to distinguish between artificial and natural nulling. If we assume that the intention of the
modulation is to serve as a beacon, one of the easiest ways to display an artificial
nature would be to have the null runs last for prime numbers of rotations. An example
histogram of null-run duration is shown in Figure~\ref{fig_nullhist}, which is extremely
unlikely to be produced due to any natural process. Another way would be to null a pulse
once after every $n$ complete rotations of the pulsar, where $n$ is a prime number. This
system might be preferable if nulling the pulsar is expensive in terms of energy, as
each null run lasts for only one complete rotation of the pulsar. In this case, a
histogram for the number of rotations between nulls would look similar to
Figure~\ref{fig_nullhist}. Any other information the ETI would like to transmit could
additionally be imposed as amplitude modulation on the non-nulled pulses.

Irrespective of whether the modulation is due to an orbiting satellite or due to a
Dysonian scaffold, during a null, if the pulsar signal is absorbed,
the temperature of the absorbing medium (modulating element) should increase.
To prevent heat build-up, the modulating element will need to shed this excess energy in
a timescale of the duration of one complete rotation of the pulsar. Looking for excess
emission with a thermal spectrum during the null phases of a pulsar would indicate such
a process in action. Considering the case of an orbiting satellite as shown in
Figure~\ref{fig_schematic}(a), assuming that the power emitted by the pulsar is given
by the spin-down luminosity $\dot{E}$ and that the modulating element is in thermal
equilibrium, the temperature of this secondary emission is
\begin{equation}
T \approx 1.1 \times 10^6 \left(\frac{r}{\rm km}\right)^{-1/2}
\left(\frac{\dot{P}}{10^{-15}}\right)^{1/4} \left(\frac{P}{\rm s}\right)^{-3/4} {\rm K},
\end{equation}
where $r$ is the orbital radius.
For a pulsar with $P = 1$ s and $\dot{P} = 10^{-15}$ s s$^{-1}$, and
taking $r = 1700$ km as derived in \S\ref{sec_modmech}, we get $T \approx 2.7 \times 10^4$ K,
which corresponds to a wavelength of approximately 107 nm, in the ultraviolet. An
excess of thermal emission that peaks in the ultraviolet during the null phases of this
pulsar, therefore, would indicate the presence of an absorbing medium.

Pulsar signals are affected by the cold plasma that makes up the interstellar
medium (ISM) in various ways \citep{ric90,cor02}. A challenging aspect of
detecting an intelligent signal within the pulsar beam is decoupling the
modulation and the effects of the ISM, particularly for pulsars in the strong
scattering regime. A simple-minded approach would be to assume that the
ETI-imposed nulling covers the entire band of radio emission from the pulsar,
whereas diffractive scintillation is frequency-dependent. Another factor that helps
discriminate between artificial modulation and scintillation is the difference in time
scales. Diffractive scintillation timescales are usually of the order of
minutes to hours \citep{col10}, which is significantly longer than any variation
due to artificial modulation, which is of the order of the pulse period.

\section{Discussion}\label{sec_disc}

The decision of using a conventional radio transmitter vis-\`{a}-vis a pulsar-based
beacon depends on the number of pulsars required to cover the entire sky, the cost of
installing modulators around those pulsars, and the lifetime of the modulating system.
Assuming that the beaming solid angle of a pulsar is about 20\% of the sky, it would
take at least five such pulsars to cover the entire sky. If the energy requirements for
sending modulating satellites to those pulsars (or building modulating Dyson shells
around those pulsars) is less than that of building and operating a perpetual,
omnidirectional, conventional transmitter of the same power, the former would be the
optimal solution, provided the lifetime of such a system is long enough.

A spacecraft that has to be inserted into orbit around another star has to accelerate
to some maximum velocity $v$ and then decelerate. To first order, the energy required for
this is twice the kinetic energy of the spacecraft from launch to achieving
maximum velocity, $E~=~m c^2 (\gamma - 1)$, where $m$ is the mass of the spacecraft
(including fuel mass) and $\gamma = 1 / \sqrt{1 - v^2/c^2}$ is the Lorentz
factor. Assuming a spacecraft mass of $10^9$ kg, travelling at a constant
velocity equal to 10\% of the speed of light, taking a time $t$ to travel a distance
equal to the minimum Earth-pulsar distance for reported values in the ATNF Pulsar
Catalogue\footnote{\url{http://www.atnf.csiro.au/people/pulsar/psrcat/}}\citep[160 pc;][]{man05}, the average energy consumption rate is approximately given by
$E/t \sim 10^{19}$ erg s$^{-1}$. Since this is much less than the energy output of a
pulsar, the cost involved in installing a few such satellites is negligible compared to
the transmission power achieved. The major factor determining feasibility is then the
lifetime of the system.

For a modulating satellite with no attitude stabilization, the lifetime depends on two
factors: (a) the radiation pressure exerted by the pulsar beam and
(b) the pressure due to particles that follow magnetic field lines impinging on the
satellite. These two effects combine to push the satellite out of its orbit. For a
modulating satellite with attitude control, the lifetime would depend on the amount of
fuel it can carry. It is conceivable that electrical energy
for attitude correction can be extracted from the pulsar beam itself, in which case,
the lifetime will be considerably longer, with an upper limit given by the radio
lifetime of the pulsar, which is about 10$^7$~yr for normal pulsars or about
10$^9$~yr for millisecond pulsars.

For a Dyson shell, the factors affecting feasibility are different. The cost incurred
in installation would be much higher than that of a modulating satellite. The lifetime
would depend on the structural properties of the scaffold, and also whether the
surrounding environment of the pulsar contains potentially destructive asteroids or
other debris.

Another factor that affects the lifetime of a satellite or a
Dyson shell around a pulsar at the distances computed in \S\ref{sec_modmech} are
radiation-induced heating and induction heating that would cause the
satellite or Dyson shell to melt and evaporate. \cite{cor08} show that an
asteroid with a radius of $\sim$~100~m at a distance of $\sim$~1000~km from a pulsar will
evaporate in less than a second. The materials used in the construction of such
systems should therefore be able to overcome heating, and would require the
development of technology not yet known to humans. For Dyson shells, another
way to reduce heating would be by having a larger radius. This would
incur a higher cost as more material would be needed for its construction. Another
way of overcoming heating would be to have a satellite orbiting the pulsar at a
safe distance and injecting material into the magnetosphere to modulate
emission.

There is one major caveat to the pulsar modulation scheme discussed in this paper, namely,
that we ignore the rate of increase of energy consumption of the extraterrestrial
civilization. If the time it takes to advance to a higher developmental stage -- one at
which the ETI can afford to build a beacon matching typical pulsar luminosity -- is less
than that for their spacecraft to reach the target pulsars, the best choice for the
civilization would be to wait. It is hard to predict rates of development, and depending
on the availability of interstellar travel technology and nearby pulsars, civilizations
may or may not choose to implement this scheme. For instance, in the event that
intelligent, technological life evolved on a planet orbiting a companion to a
pulsar in a multiple-star system, it would not only be energetically
favourable, but quicker, to implement a pulsar modulation scheme.

\section{Conclusion}\label{sec_conc}

It is reasonable to assume that energy production/consumption goes hand-in-hand
with the development of technological civilizations, as seen on Earth.
Technological civilizations should, therefore, sooner or later, embark on
large-scale energy harvesting endeavours, such as building Dyson spheres. Even
though it is unclear how inclined a civilization would be to announce their
presence explicitly using beacons, if we assume that they are so inclined,
modulating the signal of a nearby pulsar would be one of the most
energy-efficient ways of doing it. Building a Dysonian scaffold around a pulsar
would cost much less in terms of material than building a Dyson sphere at a
habitable distance around a Sun-like star, and would also be an engineering
proof-of-concept for a pre-Kardashev-Type-II civilization.

Statistical studies of pulsar emission, such as those of nulling and
pulse-to-pulse and intra-pulse intensity variation, have the potential to
discover non-natural processes in action, thereby indicating the presence of
technologically advanced civilizations in the Galaxy. Single-pulse observations
of pulsars using radio telescopes with large collecting areas will provide the
high-quality data required for this purpose.

\section*{Acknowledgements}

We thank the anonymous referee for valuable comments and suggestions, including
pointing out the possibility of injecting material into the magnetosphere to
achieve modulation. We also thank Manjari Bagchi for feedback on an earlier
version of the manuscript and Nikhil Mehta, Benetge Perera, Nipuni Palliyaguru,
and Joanna Rankin for useful discussions. A. P. V. S. and D. W. received
support from a competitive grant awarded by the John Templeton Foundation.

\end{document}